\documentclass[twocolumn,american,aps,rmp]{revtex4}
\usepackage{amsmath,amssymb}
\bibpunct{[}{]}{;}{,}{,}

\makeatletter

\providecommand{\LyX}{L\kern-.1667em\lower.25em\hbox{Y}\kern-.125emX\@}

\usepackage{pslatex,graphicx}
\usepackage[scanall]{psfrag}

\makeatother

\begin{document}

\title{\Huge State visibility in Q-bit space}

\author{A. F. Kracklauer}

\address{Bauhaus Universit\"at, Weimar, Germany}

\begin{abstract}
We study by comparison the structure of singlet type states in Q-bit space in
the light of quantum and classical paradigms. It is shown that only the
classical paradigm implies a variation in the visibility of correlation
coefficients, that has been observed in fact in experiments. We conclude that
Q-bit space in not a appropriate venue for an EPR test of quantum completeness.
\end{abstract}
\maketitle

\section{The issue}

Q-bit space, and techniques to exploit its structure, are central research topics
in modern quantum optics. Among the applications, computation, secure communication
and, of course, research on the fundamentals of physics engender intense interest. 

Still, the exact nature of quantum Q-bit space is obscure. This is a consequence
of the unresolved issues pertaining to the interpretation of Quantum Mechanics
(QM). For lack of real progress at reducing these obscure features to empirically
testable questions, however, many practical-minded researchers have lost patience
and declared the matter to be irrelevant `philosophy.'

This writer is among the exceptions who pursues this matter with the goal of
identifying features amenable to laboratory examination.\cite{1,2} There is,
in fact, one distinct point where the presently understood paradigms, if correct,
should lead to observations that can be made. If these observations differ from
what is logically expected, then the current paradigm can be questioned. Of
course, paradigm shifts, should one occur, can have no consequence for the capabilities
for exploitation of natural structure: Nature does not depend on human conceptions.

\section{States in Q-bit space}

States in Q-bit space, as is well known, can be the sum of two basis vectors
of, for example, a two dimensional space. What distinguishes this structure
from ordinary vector space structure, is that the basis vectors can be orthogonal
in the logical, rather than geometric senses. In other words, the basis vectors
can be mutually exclusive events.

This is particularly evident for states of correlated systems. The prototype is
a state for an optical Einstein-Podolsky-Rosen (EPR) experiment when the source
is chosen to emit one of two forms, i.e., either the left member is polarized
vertical while the right member is horizontal, or vise versa, that is, it is
the state of a perfectly anticorrelated pair. Now according to the current
orthodoxy, such a state for the pair as a system can be composed of a
superposition of both options, although they are mutually exclusive, e.g.:
\begin{equation}
\label{1}
\psi (1,2)=\frac{1}{\sqrt{2}}(|\uparrow >|\rightarrow >\pm |\rightarrow >|\uparrow >),
\end{equation}
in transparent notation. The usefulness of such states in the quantum
algorithms for calculating spectrographic intensities has been taken as
verification of its physical significance. But, a question that never was
resolved, is, is this a statistical state that pertains to a population of
similarly prepared systems, or does it pertain in fact to a single pair? 

The exact historical development of opinions on this matter, although intensely
fascinating, is too extensive to discuss here, so I will just state in brief
terms the general ideas. They include, that this state is ``complete,'' i.e.,
that it is taken to be a faithful symbolic rendition of an individual pair.
It is not, therefore, regarded as representing an ensemble of similarly prepared
pairs. (Of course, there are those who take exception to these assertions, Ballentine
\cite{3}, for example, but they are a minority.) 

The fact is, however, that no measurement confirms this assumption directly;
in all cases just of one of the two mutually exclusive outcomes is observed.
To accommodate this reality, an additional hypothetical input has been proposed:
``wave collapse.'' According to this hypothesis, the act of measurement itself
selects just one of the possible outcomes and destroys the other. 

These two assertions seem to have the very unscientific feature of being a logical
tautology, untestable in any experiment.

\section{Rotational invariance}

The experimental realization of the state given by Eq. (\ref{1}) is attempted
nowadays using parametric down conversion is a suitable crystal. Such crystals,
under the simulation of a single stimulus beam, emit two lower frequency outputs,
for historical reasons denoted as the `signal' and `idler,' which can be anticorrelated
with respect to polarization.\cite{4} Optimally, the crystal and frequencies
can be so chosen that these outputs overlap at two points where the conditions
implicit in Eq. (\ref{1}) are thought to be satisfied. 

One of the implicit conditions built into this expression is that it is `rotationally
invariant,' that is, that any transformation rotating axes about the wave vector
in the plane of polarization leaves it unchanged in form.\cite{5} This feature,
in combination with the two fundamental quantum assumptions mentioned above,
leads to the following imagery widely presented in textbooks on quantum mechanics.
A state comprised of the superposition of the possible outcomes, such as Eq.
(\ref{1}), is said to be in a state with no definite polarization until the
moment of measurement, when the measuring process itself causes the state to
collapse to a specific possible outcome. Further, this collapse is engendered
by either one of the two measurements, in channel 1 or 2, whichever is first.
The absolute anticorrelation of the system state dictates that an induced collapse
in either channel also induces collapse in the partner channel thereby preserving
the anticorrelation. 

Since this state is comprised of mutually exclusive outcomes, i.e., the
precollapsed state, it has a character that is denoted ``irreal.'' Moreover,
since the induced second collapse must transpire instantaneously regardless of
the separation of the measuring stations and event, this effect violates
`Einstein-locality,' according to which, all effects must be delayed with
respect to any cause by a time interval sufficient for light to propagate from
the location of the cause to the location of the effect. Thus, quantum
mechanics is said to harbor un- or irreal and nonlocal states.

The consequence of particular interest here is that the above structure implies
that the polarization of a quantum EPR state is \emph{deterministically}
anticorrelated. In other words, regardless of the direction in the plane of
polarization that a measurement is made on one arm of a single pair, if a
single `hit' is registered in that channel, then a `hit always will be
registered in, and only in, the orthogonal channel on the other arm. This is a
consequence of the fact that the precollapsed state has \emph{no} specific
polarization until measurement; so, it can acquire no specific polarization
except by agency of the polarizer filter effecting the measurement, thereby
imparting the orientation of the polarizer to the transmitted signal.
Anticorrelation then dictates that the partner component must be polarized in
the orthogonal sense. This means that \emph{no matter what angle} the polarizer
has in the channel in which the first measurement is made, if a hit is seen,
then no hit can be seen at the same angle in the other channel; a hit can occur
deterministically only in the orthogonal channel. In other words, the
probabilities of coincidences are `rotationally invariant,' as they cannot be a
function of the polarizer angle. 

This is the standard conclusion implicit in the current most widely accepted
interpretation of quantum mechanics, generally known as the ``Copenhagen'' interpretation
after the location of the laboratory of its senior proponent, Bohr.

\section{An alternative: statistical invariance}

While there are some technical exceptions, it can be said that the main alternative
to the Copenhagen interpretation is a statistical theory based on the fundamental
idea that quantum mechanics is ``incomplete,'' i.e., that it is a theory giving
only the statistical behavior of the ensemble of similarly prepared systems,
not specific information on individual systems. 

For the optical experiments testing the EPR contention (nowadays formulated
in terms of testing if a certain statistic exceeds the value \( 2 \),\cite{5})
a statistical model would be formulated on the basis of the following assumptions:

1. The source will be taken to have an axis with respect to which it emits either
one of two correlated pairs of signal pulses. One pair is comprised of a horizontally
polarized pulse in channel 1, and a vertically polarized pulse in channel 2.
The other has exchanged polarizations. 

2. The detectors in both channels will be taken to be simple polarizer filters,
i.e, they are devices that obey Malus' Law, according to which the intensity
of the passed pulse is proportional to \( \cos ^{2}\theta  \), where \( \theta  \)
is the angle the polarizer axis makes with the source axis.

3. Correlation coefficients are computed using the standard definitions for
classical signals; i.e.,
\begin{equation}
\label{2}
P(1,2)=\frac{<1|<2||2>|1>}{<1|1><2|2>},
\end{equation}
where the notation \( |1> \) is to be read as \textbar{}\( \theta _{1}> \), and
\( \theta  \) is the angle between the polarizer axis and the source axis, so
that, \( P(1,2) \) is then a function of the angles in each channel.\cite{6} It
is the coincidence intensity measured given these two angles. There are two
possibilities (vertical or horizontal; yes or no) for two channels, so that by
combination, there are four such expressions. These are measurable quantities.
The data can be collected simply by counting the number of times the signals of
the specified type are simultaneously seen in both channels as a function of
the polarizer axis angles in the respective channels; this is effected in the
laboratory by using `coincidence circuitry.'

Finally, in textbook analysis, one considers the system correlation defined
by
\begin{equation}
\label{3}
\chi (1,2)=\frac{P(v,v)-P(v,h)-P(h,v)+P(h,h)}{P(v,v)+P(v,h)+P(h,v)+P(h,h)}.
\end{equation}
This additional quantity, which pertains to the ensemble and \emph{is}
rotational invariant here too, is not measurable, but is calculated with the
individual terms. Thus, an empirical evaluation of the issue is made more
directly by determining the \( P(i,j) \) as a function of the angles. For the
nonquantum case, this is an application of Malus' Law, and so can be calculated
from first principles. 

\begin{figure*}

\psfrag{Th1=0}{$\theta_1=0$}

\psfrag{Th1=pi/8}{$\theta_1=\pi/8$}

\psfrag{Th1=3pi/16}{$\theta_1=3\pi/16$}

\psfrag{Th1=pi/4}{$\theta_1=\pi/4$}

\psfrag{Th1=3pi/8}{$\theta_1=3\pi/8$}

\psfrag{Th1=pi/2}{$\theta_1=\pi/2$}

\psfrag{P12}{$P(1,2)$}

\psfrag{Th2}{$\theta_2$}

\psfrag{Correlations, P12, as functions of polarizer angles, Th12}{{Correlations, $P(1,2)$, as functions of polarizer angles, $\theta_{1,2}$}}

\psfragscanon

\includegraphics[width=1.1\columnwidth]{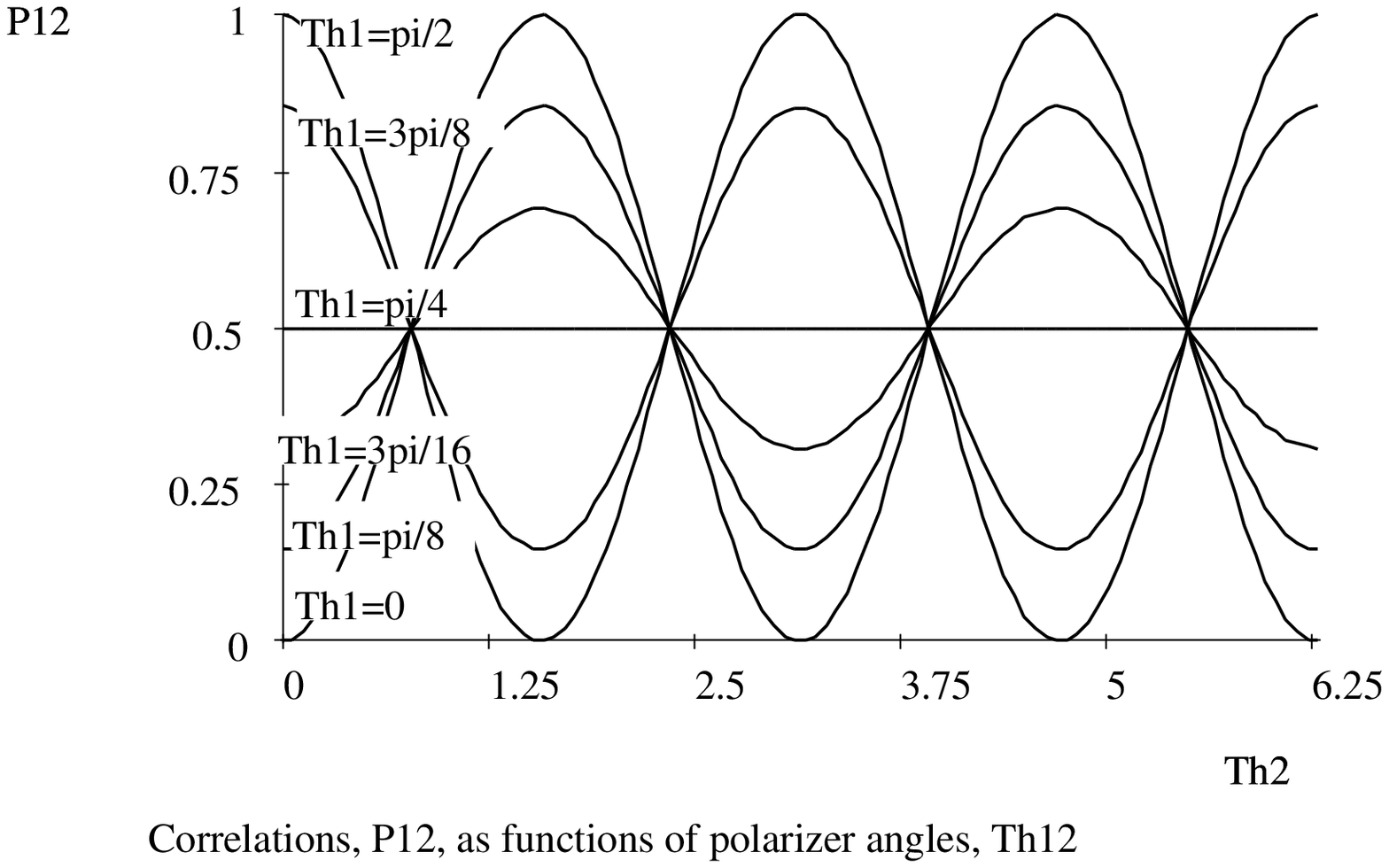}

\caption{This graph shows the calculated dependance of the coincidence
probabilitites as functions of the measurement polarizer filters with respect to
the axis of the PDC crystal axis.  Observation of this variation is
emperical support for the non quantum model of optical EPR-type experiments.}

\end{figure*}
\vspace{0.3cm}

Fig. 1 shows the result of this calculation.What it exhibits, is that the \(
P(i,j) \) are \emph{not} rotationally invariant. In particular, as the angle in
the channel in which the `first' measurement is made increases, the visibility
of the variation in the coincidence count as a function of the `second'
measurement angle changes, with a minimum of zero at \( \theta _{1}=\pi /4 \). 

Variations of exactly this character have been observed in experimental data
in fact.\cite{7}

\section{Discussion}

The quantum paradigm, on the other hand, should lead to deterministic rotational
invariance, i.e., for each individual pair.

This difference might seem to constitute an EPR test to decide the issue of
whether nature is fundamentally `real' and `local' or not. However, this
surmise is not justified by cause of another technical error in the customary
analysis. It is this: EPR proposed a \emph{Gedankenexperiment} in phase space
spanned by the operators for position and momentum. These two operators do not
commute because of Heisenberg uncertainty. For lack of practicality, the
experiment proposed by EPR could not be realized, so that alternatives were
sought. Bohm proposed changing venues to Q-bit space, as there the structure is
also noncommutative. However, the reason for noncommutativity in Q-bit space
(in partcular in its optical realization as polarization space) is not due to
Heisenberg uncertainty, but to geometry.\cite{8} This is clear as soon as one
recalls that the structure of polarization space was fully worked out by Stokes
already in 1852, many decades before the need for quantum mechanics was
appreciated. Alternately, recall that the structure of Q-bit space is encoded
in the group \( SU(2) \). This group is isomorphic to \( SO(3) \), the group of
rotations and inversions in Euclidian \( 3 \)-space. All of this structure is
obviously geometrical and has no relation whatsoever to special quantum
dynamical features such as Heisenberg uncertainty. The main conclusion here is:
Q-bit space is not an appropriate venue for EPR tests. 

Moreover, even ignoring the issue of whether Q-bit space is fundamentally
``quantum'' or not, variation of the visibility of the correlations seems to be
in conflict with the rotational invariance of singlet type states used to test
the EPR contention. This is, therefore, additional strong support for a
statistical interpretation of Quantum Mechanics.

\section*{Note}

Preprints of refs. \cite{1,2,8} can be downloaded from 
\texttt{http://www.nonloco-physics.000freehosting.com}.


\begin{thebibliography}{1}
\bibitem{1}Kracklauer, A. F., \emph{Optics and Spectroscopy,} (to appear;
quant-phy/0602080).
\bibitem{2}Kracklauer, A. F., \emph{J. Opt. B: Quantum Semiclass. Opt.,} \textbf{4}, S121
(2002).
\bibitem{3}Ballentine, L. E., `Quantum Mechanics,' (World Scientific, Singapore, 1998).
\bibitem{4}Boumeester, N., Weinfurter, H., and Zeilinger, A., in `The physics of quantum
information,' Ekert, A. and Zeilinger, A. (eds.) (Springer, Berlin, 2000). pp.
49-132.
\bibitem{5}Aspect, A., in `Quantum {[}Un{]}speakables,' Bartlmann, R. A. and Zeilinger,
A. (eds.) (Springer, Berlin, 2002) pp. 119-154.
\bibitem{6}Madel, L. and Wolf, E., `Optical coherence and quantum optics,' (Cambridge University
Press, Cambridge, 1995), Chapter 4.
\bibitem{7}Walther, T., \emph{private communication.} However, see, e.g.: Walther, T. and
Fry, E., \emph{J. Opt. B: Quantum Semiclass. Opt.,} \textbf{4}, S376 (2004),
for a description of this researcher's involvement in this issue.
\bibitem{8}Kracklauer, A. F., \emph{J. Opt. B: Quantum Semiclass. Opt.,} \textbf{6}, S544
(2004).
\end{thebibliography}
\end{document}